\RequirePackage{fix-cm}
\documentclass[smallextended]{svjour3}       
\smartqed  
\usepackage{graphicx}
\usepackage{caption} 
\usepackage{url}
\usepackage{multicol} 
\usepackage{longtable}
\usepackage{multirow}
\usepackage{float}
\usepackage[]{mdframed}
\usepackage{comment}
\usepackage{dirtytalk}
\usepackage{booktabs}
\usepackage{subcaption}  
\usepackage{tcolorbox}
\tcbuselibrary{skins}
\usepackage{natbib}
\usepackage{csquotes}
\usepackage[colorlinks=true, linkcolor=blue, citecolor=blue, urlcolor=blue]{hyperref}

\begin{document}

\title{Inclusive Employment Pathways: Career Success Factors for Autistic Individuals in Software Engineering
}

\titlerunning{Career Success Factors for Autistic Individuals in SE}        

\author{Orvila Sarker         \and
        Mona Jamshaid \and M. Ali Babar 
}


\institute{Orvila Sarker, Mona Jamshaid, Ali Babar \at
              CREST - Centre for Research on Engineering Software Technologies, School of Computer and Mathematical Sciences, The University of Adelaide \\
              \email{firstname.lastname@adelaide.edu.au}           
           \and
           Mona Jamshaid \at
              Green Pearls Consulting Pty Ltd\\
}

\date{Received: date / Accepted: date}

\maketitle

\begin{abstract}
Research has highlighted the valuable contributions of autistic individuals in the Information and Communication Technology (ICT) sector, particularly in areas such as software development, testing, and cybersecurity. Their strengths in information processing, attention to detail, innovative thinking, and commitment to high-quality outcomes in the ICT domain are well-documented. However, despite their potential, autistic individuals often face barriers in Software Engineering (SE) roles due to a lack of personalised tools, complex work environments, non-inclusive recruitment practices, limited co-worker support, challenging social dynamics and so on. Motivated by the ethical framework of the neurodiversity movement and the success of pioneering initiatives like the Dandelion program, corporate Diversity, Equity, and Inclusion (DEI) in the ICT sector has increasingly focused on autistic talent. This movement fundamentally reframes challenges not as individual deficits but as failures of environments designed for a neurotypical majority. Despite this progress, a critical gap persists: there is no integrated, evidence-based synthesis of knowledge reporting the full pathway from software engineering education through to sustainable workplace inclusion. To address this, we conducted a Systematic Review of 30 studies and identified 18 success factors grouped into four thematic categories: (1) Software Engineering Education, (2) Career and Employment Training, (3) Work Environment, and (4) Tools and Assistive Technologies. Our findings offer evidence-based recommendations for educational institutions, employers, organisations, and tool developers to enhance the inclusion of autistic individuals in SE. These include strategies for inclusive meeting and collaboration practices, accessible and structured work environments, clear role and responsibility definitions, and the provision of tailored workplace accommodations. Our findings also report the accessible and inclusive strategies for autistic SE students to help them succeed as SE professionals.


\end{abstract}

\section{Introduction}


Autism Spectrum Condition (ASC) is a lifelong neurodevelopmental condition marked by core differences in social communication, social interaction, behaviours such as insistence on sameness, unusual and varying intensity of interest or focus, as well as atypical sensory processing of the environmental stimuli  \citep{APA2013}.

It has been reported that individuals with autism possess certain faculties and cognitive styles that appear to be well aligned with the demands of careers in ICT, particularly software development.  For example, individuals with ASC are known to have exceptional abilities in information processing,  detail orientation  \citep{stuurman2019autism}, out-of-the-box thinking, and strong commitment to high-quality work \citep{begel2021remote}. Moreover, Autistic individuals usually possess strong analytical skills \citep{Wei2014}. Autistic individuals excel in software testing due to their exceptional detail-oriented focus.  Their ability to detect even minor mistakes surpasses that of their neurotypical counterparts, making them highly valuable assets in quality assurance processes \citep{haanappel2010software,stuurman2019autism}. The repetitive and precise nature of software development is inherently appealing to many autistic students \citep{annabi2017s}.  Furthermore, the consistent and logical structure of computer programming provides a predictable and rule-based environment, another factor attracting autistic individuals to the field \citep{ribu2010teaching}. Additionally, the controlled software development environment offers autistic individuals a predictable work setting with immediate feedback and minimal surprises, which can be advantageous  \citep{elshahawy2020developing}.

Employment is considered to be a marker of successful transition to adulthood and supports financial independence and psychological health. Technology companies like SAP and Microsoft have already initiated opportunities to hire computer professionals with ASC, both as a matter of social justice and to take advantage of affinities between the profile of some individuals with ASC and the job requirements of the technology industry \footnote{\url{https://www.microsoft.com/en-us/diversity/inside-microsoft/cross-disability/hiring}, \url{https://jobs.sap.com/content/Autism-at-Work/?locale=en_US}}. Research suggests that initiatives aimed at ensuring accessible software engineering education, programming tools, and inclusive work environments yield positive outcomes \citep{moster2022can,ko2017computing,zubair2023designing}. However, existing educational and employment training programs, which are essential for acquiring the knowledge necessary for software engineering roles, are plagued by accessibility issues \citep{fernandez2013mobile,eiselt2018integrating,taylor2019sex,booth2016autism,nicholas2017research}. Furthermore, employees with ASC frequently encounter challenges stemming from the lack of an inclusive work environment within the software industry. Characteristics of people with autism are that they have some limitations in the area of social skills and communicative abilities, implying that they need some additional support, guidance, and clear patterns \citep{haanappel2010software}.

To the best of our knowledge, there is currently a scarcity of research providing a comprehensive list of factors that contribute to the employment success of individuals within the software engineering field. A recent study \citep{marquez2024inclusion} discusses a high-level overview of solutions proposed in the literature to include individuals with ASC in software engineering with no rigorous discussion on improving software engineering education, career training programs, and educational tools for children with ASC. They grouped the existing solutions for the inclusion of individuals in software engineering. For example, one of the groups mentioned in their study is \enquote{more inclusive work-spaces} with little information on the actionable recommendations proposed in literature such as what can be done to set up a distraction-free work environment, how the official meeting experiences be improved, what work-related facilities and allowance can be provided to overcome the challenges faced by the employees with ASC.  To address this gap, in this study, we have performed a systematic review of 30 studies to identify the critical factors that are considered beneficial to the success of autistic individuals in the Software Engineering profession. We have classified the identified factors into four main categories: software engineering education, career and employment training, work environment and tools and assistive technologies. This study aims to benefit both educational institutions and organisations by contributing to the growing discourse on creating neurodiversity-inclusive workplaces, with a specific focus on autistic software engineers and students. This review has enabled us to identify the factors that can help organisations to successfully train and accommodate neurodivergent individuals in their SE teams. Our findings also underscore the importance of implementing accessible and inclusive strategies to support autistic SE students in successfully transitioning into professional roles within the field.

\section{Background and Motivations}
\subsection{Potential Barriers in SE Education}

Research shows that individualised and accessible software engineering education would help autistic students to get and thrive in software engineering jobs \citep{moster2022can,ko2017computing, zubair2023designing}. For example, taking part in computer programming and coding projects can help improve autistic students' communication, collaboration, and teamwork skills. Programming activities also help to develop their problem-solving skills and improve STEM learning \citep{zubair2023designing}. Coding camps can help to foster students’ interest in programming and future development jobs by providing mentoring and exposure to computing \citep{moster2022can,ko2017computing}. Unfortunately, in current software engineering education, students often face difficulties with language comprehension and executive functioning \citep{stuurman2019autism}. For instance, Code.org, a programming tool for children \citep{du2018hour}, has a complicated interface unsuitable for children with ASC \citep{fernandez2013mobile}. Again, students with ASC often struggle to initiate and maintain interactions with others while generally engaging in programming exercises on the computer \citep{eiselt2018integrating}. Students with ASC often find it challenging to adapt to university instruction due to the reduction in structure and support employed in secondary school education. University faculty members also find it difficult to connect with and teach autistic students due to a lack of knowledge and support from specialists \citep{begel2021remote}. As computer science is integrated into K-12 inclusive schools, research-based findings should inform the development of curriculum and teaching practices to ensure all students can access the content, including children with exceptionalities (formerly special needs) \citep{gribble2017cracking}. Therefore, it is essential to identify the necessary modifications to enhance software engineering education and to develop accessible programming tools to facilitate the transition of autistic individuals into the software engineering industry.   

\subsection{Challenges of Getting Hired in SE Sectors}
Autistic individuals have the desire to work but face difficulties in finding employment and maintaining employment due to the severe challenges of the work environment \citep{taylor2019sex}. Barriers to employment for autistic individuals have been attributed to multiple factors \citep{taylor2019sex,booth2016autism,nicholas2017research}. Amongst these, the job interview poses an initial barrier to employment in which strong verbal communication and overall demeanour, as expected from neurotypical (NT) individuals, are considered necessary traits for a qualifying candidate \citep{austin2017neurodiversity}. Despite good performances by a candidate for employment, to an employer who is uninformed of their skills, a candidate's performance during interviews might limit their chances of employment \citep{wehman2017effects}.  Even after successfully navigating traditional job interviews, complex work environments, including often challenging social dynamics, varied communication requirements, and the need for flexibility, it is not easy to maintain employment if appropriate support is not available \citep{baldwin2014employment, richards2012examining,roux2015national,shattuck2012postsecondary}. 

\subsection{Challenges of Thriving in SE Jobs}

Despite the potential contributions individuals with ASC can offer to software engineering roles, they frequently encounter obstacles due to the lack of inclusive workplace environments within the software industry \citep{taylor2019sex}. This includes differences or deficits in social behaviour and communication in the workplace; co-occurring conditions, such as intellectual disability or attention-deficit hyperactivity disorder (ADHD); and workplace discrimination of autistic individuals by both employers and co-workers \citep{taylor2019sex,booth2016autism,nicholas2017research}. Employees with ASC often get affected by factors associated with job tasks, such as learning new processes, and frustration associated with ongoing problems with the computer network, which prevents them from completing their tasks. Individual factors, including time management, organisation problems, and maintaining attention, can pose challenges as well. In contrast, co-workers and support staff focused on social difficulties and distractibility as the most significant barrier to trainees’ success at work \citep{baldwin2014employment, richards2012examining,roux2015national,shattuck2012postsecondary}. Other challenges include being distracted by phones, not managing work-related stress, and idiosyncratic communication styles which could be perceived as blunt or frank; these challenges need to be overcome to be successful at work \citep{hedley2018transition}. There are also challenges with various types of communications, including face-to-face conversations, phone calls, and even e-mail (particularly interpreting emotion or nuance in e-mail) \citep{morris2015understanding}. Research has highlighted that modifying the work environment and fostering inclusive organisational structures can facilitate meaningful employment for autistic individuals {\citep{hayward2019autism}.

The above examples emphasise the critical importance of identifying and devising the specific requirements and preferences of individuals with ASC for their effective integration into software engineering roles. From the perspective of neurodiversity, autistic individuals are not inherently impaired by their condition; rather, they face challenges due to environmental factors, such as insufficient awareness and understanding of autistic traits. This paradigm advocates for accommodating neurodivergent individuals' needs and celebrating their unique strengths, rather than pathologising their differences. 

\section{Research Methodology}
To report the employment success factors of people with ASC in software engineering discussed in the literature, we conducted a systematic review by following the recommended practice in software engineering \citep{kitchenham2007guidelines}. The research method was
conducted by the first author,
with suggestions and feedback from other authors. Figure \ref{Study selection Stages} demonstrates the stages of the primary study selection process and the number of studies after performing each stage.

\begin{figure} 
\centering
\includegraphics[trim=530 50 75 25,clip, scale = 0.5]{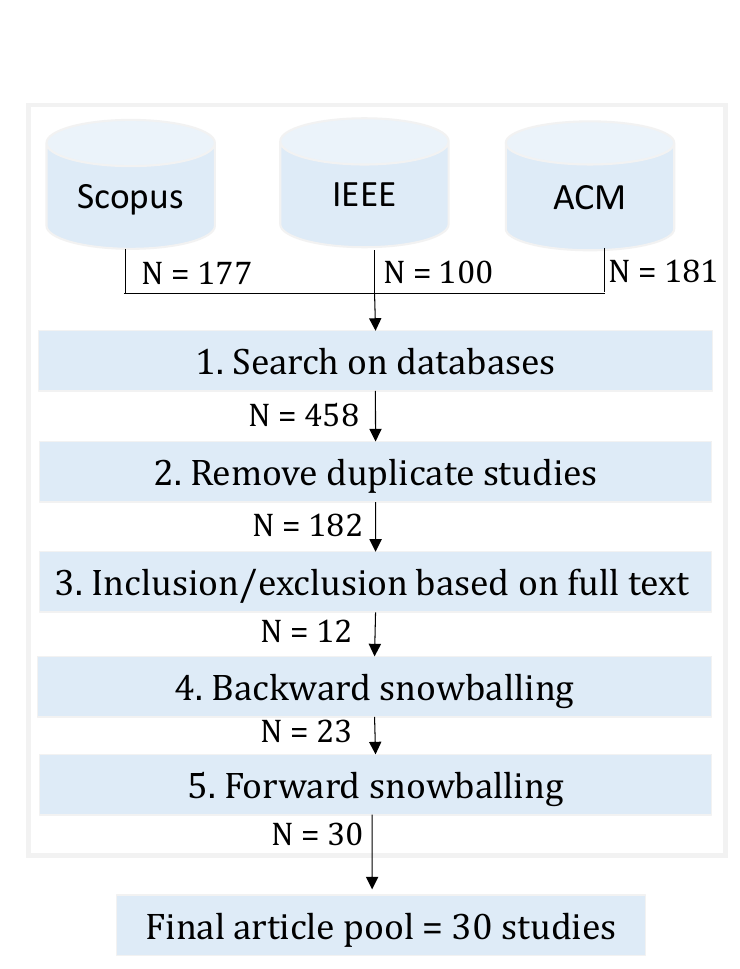}
 \caption{Study Selection Stages}
 \label{Study selection Stages}
 \end{figure}
\subsection{Search Strategy}
We searched for relevant studies in February 2024 from three academic databases - Scopus, IEEE Xplore, and ACM Digital Library. Scopus maintains a comprehensive record of a substantial number of journals and conference papers \citep{sarker2024multi}.  We included IEEE xplore and ACM digital library as these are the most frequently used databases in software engineering studies \citep{lin2020software}. To minimise the potential noise in our results, we did not consider additional databases, aligning with the approach commonly adopted in existing systematic literature reviews (SLRs) in software engineering \citep{croft2022data}.

We refined our search string based on a pilot study of 10 randomly selected studies. Initially, we used the primary term \say{\textit{austis*}} consistently and combined it with alternative terms that reflected our study’s focus. These alternative terms included \say{\textit{software develop*}}, \say{\textit{computer programming}} and \say{\textit{software engineer*}}.

In our pilot study, we noticed that some studies used the term \textit{neurodiverse}} in the title, abstract, and keywords to refer to people with autism and other neurodevelopmental conditions (e.g., \citep{tang2021understanding}). To capture those studies, we used the primary term \say{\textit{neurodivers*}} in our search string. Again, we expected that the search string \say{\textit{coding}} or \say{\textit{code}} would be suitable to choose, however, during the pilot study, we observed that even when we were restricting our search to the field of computer science, irrelevant studies from the field of Biochemistry, Genetics, and Molecular Biology were displayed (e.g., \citep{fan2023deepasdpred}). Instead of using the search string \say{\textit{programming}}, the term \say{\textit{computer programming}} provided us with more relevant results related to the scope of our study. Our final refined search string is shown below:

\begin{tcolorbox}[blanker, borderline west={1mm}{-2mm}{red}, interior engine=spartan, colback=gray!10]
\textbf{Search string:} (\textit{neurodivers*} \textbf{OR} \textit{autis*}) \textbf{AND} (\say{\textit{software develop*}} \textbf{OR} \say{\textit{computer programming}} \textbf{OR} \say{\textit{software engineer*}})
\end{tcolorbox}

\subsection{Study Selection}

\begin{table}
\centering
\caption{The inclusion and exclusion criteria}
\label{Description of inclusion and exclusion criteria}
\small
\resizebox{\columnwidth}{!}{%
\begin{tabular}{p{\textwidth}}
\\ \toprule
\textbf{Inclusion criteria}
\\ \midrule


\textbf{I1.}Studies that focus on inclusive and accessible workplaces for autistic software professionals. \\

\textbf{I2.} Studies that focus on skill development for employment in the software industry. \\  

\textbf{I3.} Studies focusing on inclusive and accessible software engineering education for autistic students. \\ 


\midrule
\textbf{Exclusion criteria}
\\ \midrule

\textbf{E1.} Studies focusing on the software technologies or applications to support the treatment of autistic people. \\

\textbf{E2.} Studies focusing on the software technologies or applications to identify the signs of autism. \\

\textbf{E3.} Studies on pedagogical techniques for autistic children, without an emphasis on software engineering education.

\textbf{E4.} Studies on the development or usability issues of assistive technologies to support autistic people with their daily routine (e.g., to understand instruction, improve the level of compliance, and communication aid).


\textbf{E5.} Studies that are not journal or conference articles.\\

\textbf{E6.} The study is a literature survey or review.\\

\textbf{E7.} Full text of the study is not available.\\

\textbf{E8.} Studies not written in English.\\

 \bottomrule
\end{tabular}
}
\normalsize
\end{table}%

\subsubsection{Inclusion/Exclusion Criteria}
A search was conducted on Scopus, IEEE xplore, and ACM digital library on the 28\textsuperscript{th} of February 2024, yielding a total of 458 articles, including 177 studies from Scopus, 100 studies from IEEE xplore, and 171 studies from ACM digital library. No filters (e.g., time period, publication type, or venue) were applied during the search process to ensure comprehensive coverage and capture of all relevant articles. Subsequently, inclusion and exclusion criteria, as outlined in Table \ref{Description of inclusion and exclusion criteria}, were employed to eliminate articles that were not relevant to the scope of this study.

Our study includes research that centres on the inclusion and accessibility of workplaces for autistic professionals in the software industry, highlighting how these environments can be adapted to support neurodiverse employees. It also encompasses studies that explore skill development initiatives aimed at enhancing employment opportunities for autistic individuals in the software field, including training programs, mentorship, and other educational interventions designed to prepare them for roles in software engineering. Additionally, the study includes research that examines inclusive and accessible educational practices within software engineering programs, specifically tailored to meet the needs of autistic students. 

We excluded studies centred on software technologies aimed at treating autism or detecting its signs and studies on pedagogical techniques for autistic children unless they specifically address software engineering education, ensuring a targeted investigation. We removed studies on assistive technologies that aid daily routines, as these do not directly contribute to the understanding of software engineering practices.  We filtered out non-journal or non-conference articles, focusing the research on peer-reviewed, credible sources, literature surveys, or reviews, favouring primary research studies that provide novel findings and the studies whose full text is unavailable or not written in English. Table \ref{Description of inclusion and exclusion criteria} displays the inclusion and exclusion criteria employed in this study.

\subsubsection{Snowballing}

Our designed search string may not be sufficient in identifying all relevant studies, as the search string may not capture studies with obscure phrasing, and the selected digital libraries do not comprehensively cover all peer-reviewed literature \citep{wohlin2014guidelines}. Therefore, following the initial study selection, we employed manual search techniques, specifically forward and backward snowballing in May 2024 and in June 2025, to retrieve additional relevant studies that were not available in our chosen digital libraries or detected by the automated search. Forward snowballing identifies further relevant studies by examining papers that cite the included studies, while backward snowballing involves reviewing the reference lists of the included studies \citep{wohlin2014guidelines}. The studies identified through snowballing were subjected to the same inclusion/exclusion criteria as the initial set. Through this process, we identified an additional 12 studies through backward snowballing and 7 studies through forward snowballing for the final analysis.

The final pool of articles comprised 30 studies, including 11 studies that passed the initial selection process and 19 additional studies identified through the snowballing technique. A complete list of the studies is provided in the Appendix \footnote{\url{https://github.com/OrvilaSarker/Neurodiversity_SR_Appendix}}.

\subsection{Data Analysis}
The raw data collected from the primary studies, aligned with the research questions guiding the investigation, were unstructured and encompassed a broad spectrum of information that posed interpretative challenges. To develop a comprehensive understanding of the data's significance and contextual relevance, and to identify recurring patterns that could aid in addressing the research questions, we employed a thematic analysis. This approach facilitates the systematic and rigorous examination of unstructured data, allowing for the exploration of its intricacies and nuances \citep{sarker2024multi}. Consequently, thematic analysis offers valuable insights into the complexities inherent in unstructured data.

We followed the thematic analysis framework outlined by Braun and Clarke \citep{braun2006using}. The analysis was conducted using NVivo, a qualitative data analysis software \citep{braun2006using}. The extracted data, initially stored in an Excel spreadsheet, was imported into NVivo, where we performed open coding. Open coding involves segmenting the data into smaller units and assigning labels (or codes) to each segment \citep{sbaraini2011grounded}. This process was iterative, with codes generated during the initial phase being revised and updated in subsequent stages.

\begin{figure} 
\centering
\includegraphics[trim=0 235 175 4,clip, width = \textwidth]{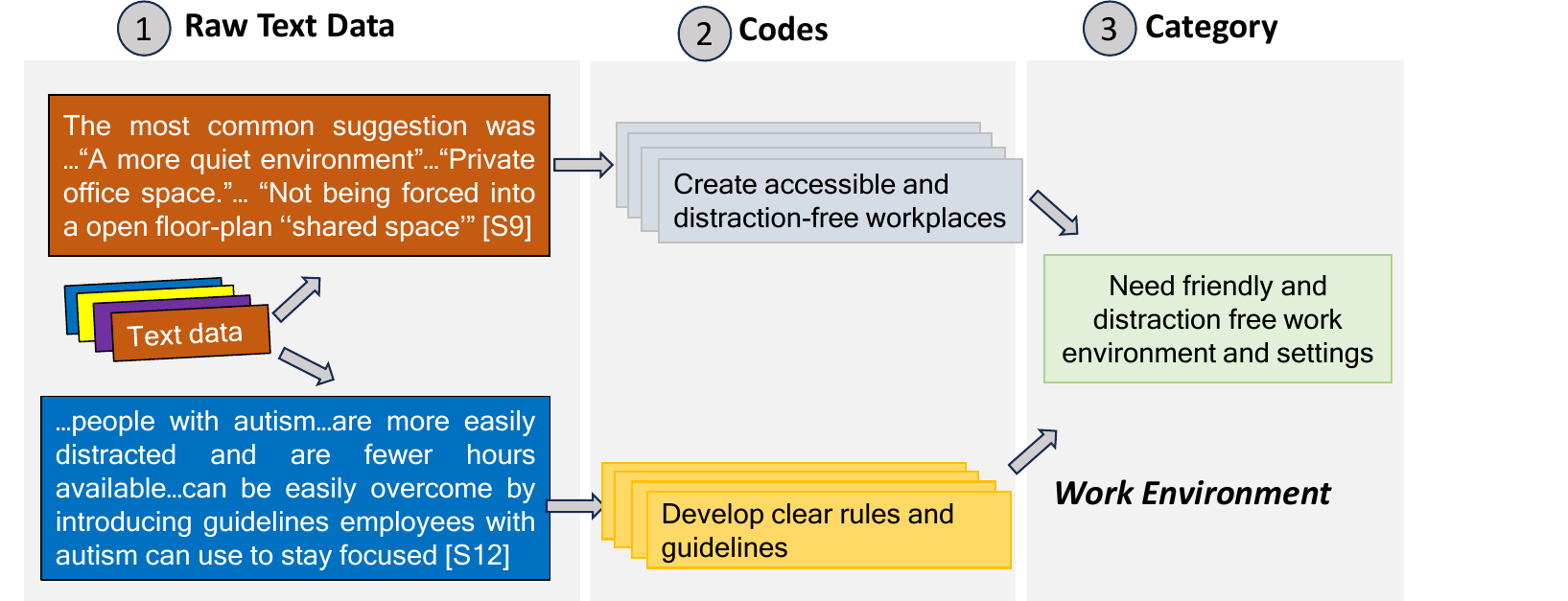}
 \caption{Data Analysis}
 \label{Data Analysis}
 \end{figure}

To familiarise ourselves with the data patterns, the first author conducted a pilot data extraction on a randomly selected subset of five academic studies. The first author thoroughly examined the codes and grouped them into broader themes, based on their similarities, using NVivo’s multi-layered structure. The second and third authors acted as validators, providing feedback and suggestions that were incorporated into the ongoing data analysis process. Figure \ref{Data Analysis} shows our data analysis process.

\section{Overview of the Selected Studies}

\begin{figure}[htbp]
    \centering
    \begin{subfigure}[t]{0.50\textwidth}
        \includegraphics[width=\linewidth, trim=10 0 10 0, clip]{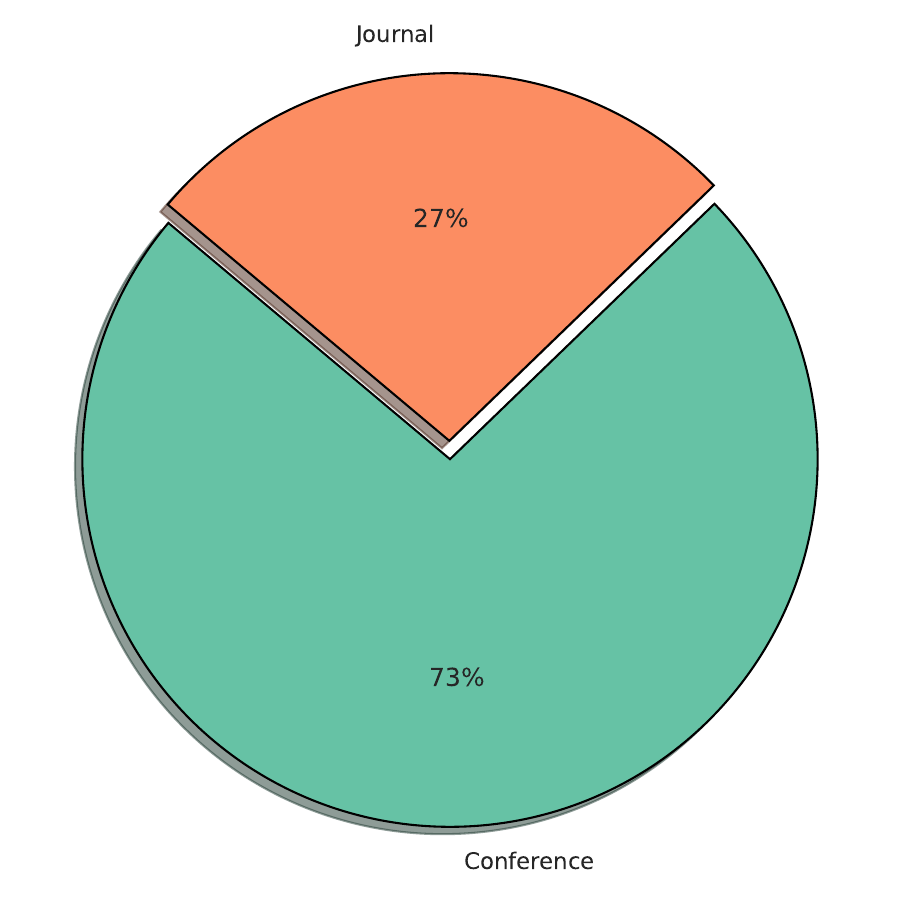}
        \caption{Publication types of the selected studies}
    \end{subfigure}
    \hfill
    \begin{subfigure}[t]{1\textwidth}
        \includegraphics[width=\linewidth, trim=50 0 60 0, clip]{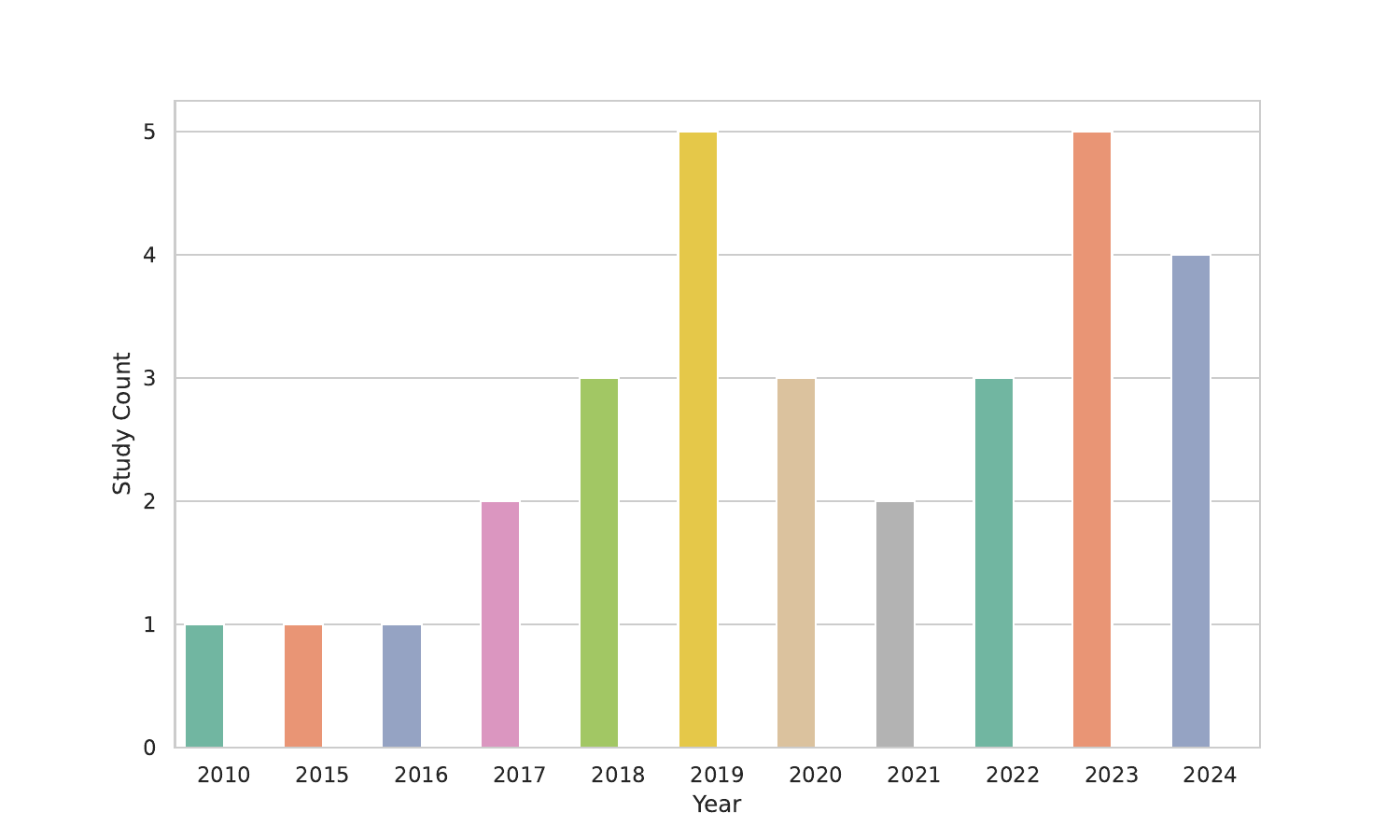}
        \caption{Study count for each year}
    \end{subfigure}
    \hfill
    \begin{subfigure}[t]{1\textwidth}
        \includegraphics[width=\linewidth, trim=0 0 0 0, clip]{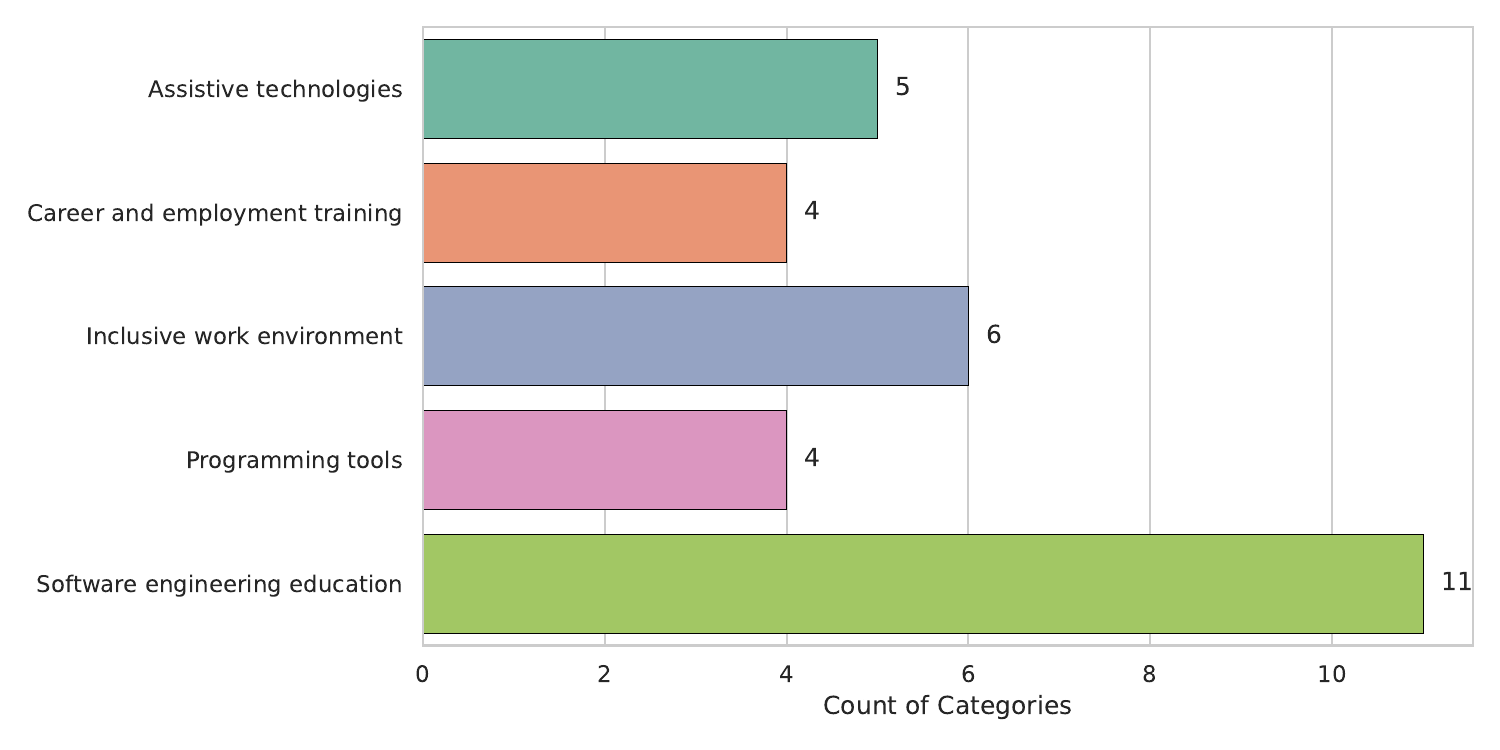}
        \caption{Main topics of the studies}
    \end{subfigure}
    \caption{Demographic details of the primary studies}
    \label{fig:three-images}
\label{Demographic details of the primary studies}
\end{figure}

Figure \ref{Demographic details of the primary studies} depicts the demographic information of our primary studies. Our study pool includes 21 conference papers, including ICSE, CHI, CSCW, and ASSETS venues and 9 journal papers including Computers \& Education, IEEE Software, ACM Human-Computer Interaction, The Journal of Systems and Software, ACM Transactions on Accessible Computing and Journal of Management \& Organisation (Figure \ref{Demographic details of the primary studies} (a)). Our studies published are within the range of 2010 to 2025 (data for 2025 is incomplete as we conducted our second stage forward snowballing in early 2025), and most of the studies were published in 2018 and 2024 (Figure \ref{Demographic details of the primary studies} (b)). Our study pool consists of 5 studies on assistive technologies, 4 studies on career and employment training, 6 studies on the inclusive work environment, 4 studies on Programming tools, and 11 studies on software engineering education (Figure \ref{Demographic details of the primary studies} (c)).





\section{Findings of the Study}
\subsection{Software Engineering Education}
\subsubsection{Provide individualised support to the students with ASC}
Studies found that autistic students perform better if provided individualised support [S4]. Autistic students face challenges in performing abstract and inferential tasks such as writing passages, passage comprehension, and solving applied math problems [S4]. Certain pedagogy is found effective for autistic children, such as unplugged computing, physical computing, and implementing visual learning [S24]. To help students perform better, it is important to provide them with inclusive educational settings. Therefore, several studies have suggested adjusting the education settings according to autistic students' needs, such as additional classes for autistic students [S4], providing visual support for specific lessons [S24], and content-specific support [S24].
For programming classes and coding camps, providing autistic students computers that are pre-installed with all of their requirements, for example, having VS code software and Zoom application installed, having more than one monitor available to be able to switch between windows, etc. Having a pre-built environment can solve the requirements for addressing discrepancies among operating systems and different UI needs [S2]. For coding camps, instructors should survey students to understand what technical features they plan to implement and modify the lecture materials accordingly. Knowing the required system features would help the students save time to stay on track to implement the desired task [S24].

For providing individualised support, an educator with specialised knowledge, including characteristics of disabilities, accommodations for learning, and strategies that promote engagement, is recommended [S3]. A specialised educator can help to avoid student over-stressing, clarify instructions [S3], ensure that student-specific support is in place [S24], and maintain ordered communication [S3]. 

\subsubsection{Create structured teaching instructions}
Summarising and generalising is sometimes challenging for autistic students [S4]. Therefore, teaching instructions should be simple and well-structured with minimum text, more examples, a simple presentation layout [S19], and more video-based instructions and tutorials [S19].  If guidelines are provided on how to perform a specific task, they should be accompanied by concrete examples [S4]. Teaching instructions need to be straightforward with simple visual layouts with clearly marked
headings, example illustrations, and checklists [S19]. Step-by-step instructions would allow students with ASC to follow the lecture at their own pace without interrupting the instructor and peers. Scheduled breaks in between lectures are recommended to help the students socialise with their classmates [S2].

\subsubsection{Educate the educators and peers}
Building teachers' knowledge and awareness regarding autism can provide autistic students with better access to software engineering education. Research has demonstrated that enabling teachers on the cognitive styles and social aspects, cognitive deficiencies, and potentials of a student with autism are the keys to autistic students' success in software engineering education [S2, S24, S25].
Training programs for teachers can be implemented to effectively implement the pedagogies for students with ASC [S25]. An enlightened educator can develop effective strategies to better support the needs, for example, by investigating different ways that bring back interest and increase attention to programming. Teachers can observe and deliver course materials based on the interests of their autistic students.

\subsubsection{Offer opportunities to improve collaboration and problem-solving skills}

Arranging coding camps for autistic students fosters their interest in programming and increases employment opportunities in software development by exposing them to computing and mentoring opportunities. The coding camp's teamwork setup helps students learn time management and improve their interpersonal skills, including collaboration and communication skills, which are essential for jobs in software development [S2, S5]. Through teamwork and collaboration, students can share their ideas and review others' code. By forming a personal connection, students gain confidence. Coding camps and collaborative programming classes provide opportunities to practice teaming [S2, S5]. Gaming environments provide opportunities for autistic students to build friendships and to negotiate new social situations required in software development [S2]. To minimise the anxiety that comes from working with a new set of students, coding camps may consider keeping the same team throughout the camp [S5].

\subsubsection{Assist in future career development}

To support adults with ASC in finding and maintaining employment in the software industry, state-level and government-level employment training programs can be provided to the students to help them gain experience in real-life work environments that match their interests [S2]. Some autistic students face challenges in making friendships in face-to-face collaboration required in university education systems. Therefore, universities may arrange summer online courses to provide first-year university students with a taste of university education and an opportunity to connect with their peers [S5]. To provide career-building advice and personal career experience in STEM, expert presentations, panel discussions, and seminars can be organised [S3].

\subsection{Career and Employment Training}

\subsubsection{Help develop social and collaboration skills}

Research has shown that if students with ASC were provided internship opportunities and collaboration opportunities, their task management, social, and collaboration skills were improved [S11, S20]. Students showed their creativity and leadership skills through the exchange of creative ideas, besides technical tasks such as voluntarily helping someone with coding and software, and implementing non-functional requirements of software development, such as programming a machine to tell a joke [S20].  Teaching and learning programs should allow peer mentoring, which can help the student with ASC to learn the responsibilities of mentorship and develop leadership skills [S19]. Groupwork helps an individual with ASC to overcome individual challenges. Therefore, group activities can be designed for planning, generating code, and assembling hardware [S19].  

\small
{
\setlength{\tabcolsep}{3pt}

\begin{longtable}{p{0.27\textwidth}p{0.65\textwidth}p{0.06\textwidth}} 
\caption{Factors beneficial to enable individuals with ASC to succeed in the software industry (education \& training)} \label{table:csfs_in_PETA}\\

\toprule
\textbf{Factors} & \textbf{Key points (included papers)} & \textbf{\#} \\
\midrule
\endfirsthead

\toprule
\textbf{Factors} & \textbf{Key points (included papers)} & \textbf{\#} \\
\midrule
\endhead

\midrule
\multicolumn{3}{r}{\small\textit{(Continued on next page)}} \\
\midrule
\endfoot

\bottomrule
\endlastfoot
 \multicolumn{3}{c}{\textbf{Software Engineering Education}} \\
 \midrule
1. Provide individualized support to the students with ASC & \textbullet\ Incorporate effective pedagogy for students with ASC to easy to follow [S4, S24] \newline \textbullet\ Conduct surveys with the students to modify the lecture [S2] \newline \textbullet\ Allow extra tutoring time for students with slow progress [S3] \newline \textbullet\ Have a pre-built environment with specific computer needs for coding camp [S2]  \newline \textbullet\ Allocate resources to support students with ASC [S24, S28] \newline \textbullet\ Assign a specialized educator in the camp to satisfy a variety of teaching needs [S3] \newline \textbullet\ Arrange student-specific support [S24] \newline \textbullet\ Develop standardized criteria to guide the selection of the suitable technological resources to meet educational needs [S28] \newline \textbullet\ Provide scafolded instructions [S30] & 6 \\

2. Create structured teaching instructions & \textbullet\ Teaching instructions should be simple and well-structured [S19] \newline \textbullet\ Teaching instructions with visual aids [S19]\newline \textbullet\ Allocate instructor for each groups in coding camps [S5]\newline \textbullet\ Explicit guidelines with exercise [S4]\newline \textbullet\ Provide step-by-step guidance and create predefined activities [S2, S4] & 4 \\\cline{1-3}

3. Educate the educators, caregivers and peers & \textbullet\ Pre-recorded videos for educators to practice their sessions [S2] \newline \textbullet\ Teaching children to support their autistic peers [S4] \newline \textbullet\ Educators should develop knowledge about inclusive education and teaching methodologies [S2, S24, S25] \newline \textbullet\ Provide gender-specific training [S24] \newline \textbullet\ Promote awareness among educators and caregivers of available technological applications
to understand the quality elements underpinning their
effectiveness [S28] \newline \textbullet\ Develop guidelines to support
methodological selection and evidence-based evaluation of technological applications [S28] & 5 \\\cline{1-3}

4. Offer opportunities to improve collaboration and problem-solving skills & \textbullet\ Organise coding camp to build teamwork and communication skills [S2, S5]\newline \textbullet\ Introduce game-based programming to help develop problem-solving skills [S5]\newline \textbullet\ Let people work with the same collaborators [S5] \newline \textbullet\ Increase social translucence to make people with ASC more aware and accountable [S5] & 2 \\\cline{1-3}

5. Assist in future career development & \textbullet\ Real-world work training to help gain exposure to workplace in tech industry [S2] \newline \textbullet\ Provide summer online course to get a taste of the universities [S5] \newline \textbullet\ Arrange career seminars to provide the opportunity to know the working environment [S3]\newline \textbullet\ Provide financial support for the education of autistic children [S3] & 3 \\\cline{1-3}
\multicolumn{3}{c}{\textbf{Career and Employment Training}} \\
 \midrule
6. Help develop social and collaboration skills  & \textbullet\ Offer internships to encourage task engagement and social interaction [S11, S20] \newline \textbullet\ Allow collaborating with other developers to improve collaboration skills [S26]\newline \textbullet\ Allow group conversation to help identify problems [S19]\newline \textbullet\ Design group work to overcome an individual’s accessibility challenges [S19]\newline \textbullet\ Support peer mentoring to help reinforce the responsibilities of mentorship [S19] \newline \textbullet\ Offer learning environment to help develop social skills such as communication required to work with a team [S30] & 5 \\\cline{1-3}

7. Assign experienced trainers and job coaches   & \textbullet\ Educate the trainers about STEM tools [S23]\newline \textbullet\ Staff need to be open to diversity and flexible to adapt both the environment and process content [S20] \newline \textbullet\ Having supportive mentors to understand and address specific needs of people with ASC [S12, S22, S26] & 5 \\

8. Provide individualized support  & \textbullet\ Accommodate unique learning needs [S26]\newline \textbullet\ Accessible learning materials [S26] \newline \textbullet\ Record lectures for students with slow progress [S26] \newline \textbullet\ Allow clients to work and spend more time with the STEAM toolkit [S23]\newline \textbullet\ Include a neurodivergent person in the training and development team [S26] \newline \textbullet\ Provide necessary work equipment [S26]\newline \textbullet\ Create prompts with visibility and limit the need for large memory recall [S23] \newline \textbullet\ Encourage diverse goals and outcomes [S19] \newline \textbullet\ Support to break down problems into smaller steps [S19] \newline \textbullet\ Provide flexibility regarding workplace environment [S22] \newline \textbullet\ Emotional support from peers and supervisors [S11]\newline \textbullet\ Have training camps to provide practical training [S11, S22, S26] & 5 \\\cline{1-3}

9. Provide structured and in-advance training materials & \textbullet\ Pre-notice for upcoming tasks and opportunities to take breaks [S23] \newline \textbullet\ Provide simple, well structured and modular activities [S19]\newline \textbullet\ Reducing unstructured time, direct communication and maintaining schedules [S22]\newline \textbullet\ Provide explicit instruction [S23]\newline \textbullet\ Make the teaching materials available in advance [S26]  & 4\\\cline{1-3}

10. Develop ASC-friendly recruitment strategy  & \textbullet\ For training, provide coherent yet tangible feedback to the unsuccessful candidates [S20]\newline \textbullet\ Employer or recruiter should be aware of the neurological traits[S20]\newline \textbullet\ Include autism employment in the recruitment strategy [S20] \newline \textbullet\ Assess training programs through observational assessment of clients [S23] \newline \textbullet\ Use structured interview questions [S11]  & 3 \\

\end{longtable}
  
\normalsize
\subsubsection{Assign experienced trainers and job coaches}

Support from the organisation, training staff, and co-workers can improve the employment success of individuals with ASC in the software industry [S22]. To accommodate the needs of autistic individuals, training staff need to be well-trained and open to diversity and inclusion to ensure autistic individuals have an equitable opportunity to retain the training information [S20, S22]. It is important to train the staff members so that they are familiar with the best accessibility training practices and are capable of designing accessible materials, such as codes, presentations, and lectures [S26]. Experienced and knowledgeable staff can modify the training content and change the communication styles to reduce misunderstandings.  Coaches and trainers need to understand how to use STEAM kits to train individuals with ASC in career training organisations [S22]. Career training coaches should be capable of administering teaching materials inclusively, including navigating code using assistive technologies such as screen readers. Nonetheless, many software developers exhibit a deficiency in accessibility expertise, thereby complicating the recruitment of qualified instructors for these programs. Consequently, it is imperative to garner support from companies to enhance proficiency in accessibility teaching practices and gather continuous feedback from participants to comprehend their evolving requirements [S26].

\subsubsection{Provide individualized support}

Providing personalised support and continuously improving the learning materials is crucial to help develop skills for employment in software development, as discussed in studies [S11, S19, S22, S23, S26]. Studies recommended adding detailed descriptions of the images in the figure, slower explanations, and avoiding certain phrases such as "on the screen, I am doing this", and "click with the mouse here" [S26].  Accommodating unique needs such as allowing to keep the camera or microphone turned off and allowing to send questions on the chat in the online classes rather than talking etc [S26].  Recording lectures can positively impact the learning experience of individuals with ASC as this allows learning at their own pace. Understanding training participants' needs and requirements helps them to get a better training experience such as providing laptops with pre-installed software, wide-screen monitors [S26], dividing the project into smaller tasks -  writing down the program structure before starting coding [S19], providing flexibility for workplace settings - adjusting the lighting of the room, providing headphones to help avoid noise, etc. [S22], allowing the training participants to engage with the STEAM toolkits to help them learn skills [S23]. Study S26 suggested that including neurodiverse people in the design and development of learning materials creates an inclusive environment [S26].

\subsubsection{Provide structured and in-advance training materials}

Supplying educational materials ahead of lectures, sending pre-notices for upcoming tasks, and providing well-structured and simple instructions are advocated for various reasons [S19, S22, S23, S26]. Firstly, providing code snippets accompanied by concise explanatory files enhances the understanding of neurodivergent participants. Secondly, it reduces anxiety among participants by providing insight into the forthcoming discussion topics, a particularly beneficial aspect in lectures characterised by a slow pace. Thirdly, it enables participants to pinpoint and address accessibility concerns before the lecture session. Fourthly, it serves as a resource for students who may have missed a lecture, thereby aiding their learning process [S26]. Receiving advance notice of work assignments and opportunities for periodic breaks can assist neurodivergent individuals in making strategies to manage stress.

\subsubsection{Develop ASC-friendly recruitment strategy}

Studies suggested several modifications in the recruitment strategy to accommodate the needs of individuals with ASC disorder. Self-disclosure of a neurodivergent condition necessitates that the neurodivergent condition will not adversely affect the opportunities within or beyond the employment training program's confines. The recruitment of adults with neurodivergent conditions has not been considered as an organisational imperative; consequently, the US government has incentivised organisations to engage in the hiring of neurodivergent applicants. To mitigate the ethical hazard associated with corporations viewing disability employment solely as a means to capitalise on financial incentives, it is recommended that organisations integrate the recruitment and selection process into their overarching strategic framework [S20]. Candidates who applied for the employment training programs but remain unsuccessful should receive positive feedback to improve their employability [S20]. Employers or recruiters need to be aware of how to identify or associate with neurodivergent people to avoid making a negative inference about one’s capability [S20]. For interviewing individuals with ASC disorder, it is suggested to use structured interview questions as opposed to open questions. Structured questions are more effective for eliciting memories and personal facts. When given visual-verbal prompts (i.e., a question about a particular memory is split into: the date and time of the memory, involved people and the setting, the events and actions that took place, and the involved objects), individuals with ASC disorder can easily recall the information [S11]. The provided employment training programs can be improved through observational assessment of the clients (i.e., the neurodivergent individuals) [S23].


\subsection{Work Environment}

\subsubsection{Need a friendly and distraction-free work environment and settings}

Several studies have discussed the importance of accessible, friendly, and distraction-free office environments to help individuals with ASC disorder to stay focused and feel safe [S8, S9, S10, S12, S13, S20].  Some examples of desirable office set-ups mentioned by individuals with ASC are computer monitors at the correct height, a suitably arranged space for using the keyboard, noise-cancelling headphones, and quiet office rooms [S9]. Individuals with Autism Spectrum Disorder (ASC) are prone to distraction [S12]. The primary disadvantage of this tendency can be mitigated through the implementation of clear rules and guidelines that help employees with autism maintain focus [S12]. 

\subsubsection{Ensure structured and organised meetings and video conferences}

It is common for professionals with ASC to get distracted during online video conferences and meetings [S13].
Interrupting for a recap of missing parts of the meeting is considered socially impolite. Thus, requesting quick recaps after inadvertent moments of distraction needs to be normalised in the organisational norms around attention management [S13].  Allowing neurodivergent individuals to record meeting audio and take notes is also recommended [S8, S10]. Neurotypical colleagues should remain mindful and supportive of such re-focusing needs of their neurodivergent team members. Remote video conferencing applications can include a ‘request quick recap’ feature to support the needs of neurodivergent professionals. This functionality can also be configured to automatically deliver the last few seconds of verbal interaction in text or audio/video format. Specifically, it could offer a reference to a segment of live text transcription [S13]. Research showed that people with ASC usually feel more comfortable communicating over text messages than phone calls, video calls, and face-to-face conversations [S13, S14]. Employers should respect these preferences and communicate with employees in their preferred communication method.

The option to work remotely allows individuals to regulate sensory stimuli within their home environment, which, coupled with flexible scheduling and proximity to support resources, contributes to the well-being of individuals with ASC. Reduced meeting attendance permits these individuals to dedicate more time to their strengths, such as coding. Remote work facilitates more accessible and adaptable work routines, enabling individuals to rest and then resume work, thereby enhancing productivity throughout the day. 

\small 
\setlength{\tabcolsep}{4pt}
\renewcommand{\arraystretch}{1.1}
\begin{longtable}{p{0.27\textwidth}p{0.65\textwidth}p{0.06\textwidth}} 
 \caption{Factors beneficial to enable individuals with ASC to succeed in the software industry (work environment \& tools)}
\label{table:csfs in PETA-ASC}\\

\toprule
\textbf{Factors} & \textbf{Key points (included papers)} & \textbf{\#} \\
\midrule
\endfirsthead

\toprule
\textbf{Factors} & \textbf{Key points (included papers)} & \textbf{\#} \\
\midrule
\endhead

\midrule
\multicolumn{3}{r}{\small\textit{(Continued on next page)}} \\
\midrule
\endfoot

\bottomrule
\endlastfoot

\multicolumn{3}{c}{\textbf{Work Environment}} \\
 \midrule
11. Need friendly and distraction-free work environment and settings & \textbullet\ Create accessible and distraction-free workplaces [S9, S13] \newline \textbullet\ Friendly physical environment where ASC employees do not hesitate to ask for help [S8] \newline \textbullet\ Develop clear rules and guidelines [S12] \newline \textbullet\  Suitable physical environment such as correct monitor height, noise cancellation headphones, treadmill desk [S8, S9, S10, S12]  & 6 \\\cline{1-3}

12. Ensure structured and organized meetings and video conferences  & \textbullet\ Organise meeting questions and share meeting materials in advance [S13]\newline \textbullet\ Allow audio recording of the meeting and taking notes [S8, S10] \newline \textbullet\ Provide meeting summary to ASC employees [S8] \newline \textbullet\ Support refocusing after periods of distraction [S13]\newline \textbullet\ Allow ASC employees to process information [S10] \newline \textbullet\ Preferences to text messaging [S13, S14]\newline \textbullet\ Work from home facility with fewer meetings [S9, S10, S12, S13, S17] & 7 \\\cline{1-3}

13. Assign roles and responsibilities suitable to individuals with ASC  & \textbullet\ Clearly define the goal of the project [S10]\newline \textbullet\ Breakdown large tasks into small ones with close interest [S8, S10, S12] \newline \textbullet\ Getting autonomy to code the part of the system [S9] \newline \textbullet\ Allow to work on a solo project [S9]\newline \textbullet\ Set up realistic timelines and deadlines to reduce the stress of ASC employees [S8] \newline \textbullet\ Design roles suitable to the unique abilities of ASC people [S10]& 4 \\\cline{1-3}
    
14. Provide necessary work-related facilities and allowances & \textbullet\ Better health insurance for caring for children with ASC [S9]\newline \textbullet\ Private hotel room for business travel [S9] \newline \textbullet\ Fly in business class to avoid stress caused by the crowd [S9]& 1 \\\cline{1-3}   

15. Create awareness about ASC among neurotypical employees and employers & \textbullet\ Training  managers and peers to understand inclusivity [S8, S10] \newline \textbullet\ Disclosing ASC disorder to employers can increase social inclusion in the workplace culture [S20]\newline \textbullet\ Greater awareness and sensitivity among neurotypical employees [S9] \newline & 4 \\\cline{1-3}

\multicolumn{3}{c}{\textbf{Tools and Assistive Technologies}} \\
 \midrule

16. Leverage assistive technologies for employment training and task completion & \textbullet\ Design educational material with visual aids [S19]\newline \textbullet\ Use of assistive technologies to provide support to job training and task completion [S11, S13, S14, S19  ]\newline \textbullet\ Use virtual interactive training [S11] & 4 \\\cline{1-3}

17. Use CMC technologies to help increase attention and flexibility & \textbullet\ Create plain backgrounds to help avoid distraction [S2]\newline \textbullet\ Use different features of CMC technologies to increase the focus and attention [S14, S18] \newline \textbullet\ Minimise the video calling screen to turn the attention away [S13] & 4 \\\cline{1-3}

18. Design accessible and personlaised tools and technologies    & \textbullet\ Design simple and game-based visual programming tools [S3, S5, S6, S7, S14, S15, S16, S17, S24, S18, S19, S26, S27] \newline \textbullet\ Need interdisciplinary collaboration between therapists and software engineers to design accessible software [S28]\newline \textbullet\ Integrate diverse communication mechanisms (e.g., audio, text, narratives) in the design phases of technologies [S28] \newline \textbullet\ To promote self-perception of learning, embed visualisation of the activity results [S28] \newline \textbullet\ Design motivating, engaging interfaces [S28] \newline \textbullet\ Ensure proper maintenance support and updated information [S28] \newline \textbullet\ Provide friendly navigation and reduce content overload [S28] \newline \textbullet\ Evaluate commercial support by restricting unauthorized in-app purchases and adopting alternative revenue models that are non-disruptive and accessible [S28] \newline \textbullet\ Use of LLMs, LVMs and IOT-based sensing networks to collect behavioral and physiological data for designing personalized training content [S29] & 14 \\

\end{longtable}
   
\normalsize
Additionally, the elimination of commuting time and effort significantly benefits their mental well-being and concentration [S9, S10, S12, S13, S17].

Lack of prior knowledge about the discussion topics of the upcoming meeting causes a significant cognitive burden to employees with ASC [S13]. It is recommended in the literature to share meeting materials and agendas in advance to facilitate these individuals to easily follow the discussion during meetings when they try to re-focus on the meeting after a moment of distraction and need to get the context of the ongoing conversation.
Sharing meeting goals helps these individuals to meet the expectations of the leaders and managers and helps them not receive negative feedback, which causes stress [S13]. Studies also suggested providing a summary of important meeting decisions [S8].

The literature indicates that video conferencing tools can facilitate the integration of meeting agendas into remote meetings, enabling the tracking of agenda items in real time to offer a visual status update and reminder. Interactive meeting agendas and supplementary materials (such as notes and slides) could enhance the user experience by supporting navigation through transcripts and recordings post-meeting. For instance, clicking on an agenda item or slide would direct users to the corresponding point in the transcript or recording. These features could be particularly beneficial for neurodivergent professionals by indicating where they should focus if they miss part of the meeting for a desensitisation break or become distracted and need a recap. Additionally, an interactive agenda could assist in monitoring the meeting's pace and timing. The ‘5 minutes left reminder’ feature in Microsoft Teams aids in concluding meetings promptly. This feature could be further developed to track the timing of individual agenda items and optionally provide reminders to the meeting host and/or attendees, thereby helping to maintain structure in remote meetings [S13]. The 'hand raise' feature in video conferencing tools enables participants with ASC to signal their intention to ask a question. This functionality should remain accessible and operational even during screen sharing [S13]. Furthermore, a structured approach to question management can be employed by maintaining an orderly queue in a separate Google Doc. This document would list the names of attendees who have indicated they have questions, and individuals would be called upon in turn [S13].

\subsubsection{Assign roles and responsibilities suitable to individuals with ASC}
Individuals with ASC frequently pursue and analyse information to categorise and integrate it into specific frameworks. To optimise the distinctive cognitive characteristics of employees with ASC, tasks related to their roles must be structured and communicated with clarity. Furthermore, it is crucial to design roles that can be most effectively fulfilled by these specialised skills [S10]. Not being able to meet unrealistic deadlines causes anxiety and stress among software developers with ASC [S8]. It is suggested to set longer deadlines for software development tasks, which would help the employees avoid working extra hours to compensate for delays caused by the bad estimates. Breaking down larger tasks into smaller ones in agile software development would help in task initiation and in having a motivating sense of accomplishment after finishing the tasks. Sometimes while focusing on achieving a big goal, employees might end up getting lost and not being able to accomplish anything [S8, S10, S12]. Also, individuals should be allowed to work on solo projects and have the autonomy to decide which part of the system they want to code [S9].

\subsubsection{Provide necessary work-related facilities and allowances}

A study [S9] found that employees with ASC, whose children are also diagnosed with ASC, recognise the importance of comprehensive health insurance coverage for the care of their children. This enhanced coverage is likely to be highly valued by many employees with ASC, as their children are statistically more likely to have ASC compared to the children of neurotypical employees. Study [S9] also examined the preference of employees with Autism Spectrum Disorder (ASC) for private hotel accommodations during business travel. This preference is especially pertinent as the social interactions required during travel can be exhausting for these individuals. A private room would allow them to relax and recharge in solitude, mitigating social fatigue's effects. Additionally, other work-related social activities, such as team dinners, can be equally draining for employees with ASC. Furthermore, these individuals may opt for economy class flights during business trips due to the heightened stress associated with being near numerous people, which is exacerbated by their condition.

\subsubsection{Create awareness about ASC among neurotypical employees and employers}
Research underscores the significance of fostering social awareness among neurotypical employees and employers to cultivate an inclusive work environment [S8, S9, S10, S20]. The literature advises that managers conduct regular, non-judgmental, one-on-one discussions with software practitioners who have ASC. Peer mentors play a crucial role in establishing a culture of understanding and inclusiveness. Additionally, stigma and ignorance associated with ASC can make the workplace less welcoming for individuals with diverse cognitive and interpersonal traits. Thus, it is imperative to educate managers and peers about the characteristics of ASC. This training should be supplemented with performance-based rewards to ensure accountability for inappropriate behaviours toward individuals on the spectrum [S8, S10].

Disclosure of a neurodiverse condition allows employers to provide appropriate accommodations and ensures organisational compliance with federal regulations [S20]. Recruitment workshops, for example, should be tailored to accommodate the neurological sensitivities of participants with ASC, such as avoiding music, minimising decor, and utilising natural sunlight. Moreover, disclosure of a disability can reduce the likelihood of employees concealing their condition, thereby enhancing social inclusion within the workplace culture. Employers can also observe employees to identify behaviours indicating difficulty with tasks (e.g., during meetings) or high anxiety levels. This enables the implementation of policies that ensure equitable opportunities for all employees [S20].

Creating an environment that educates all employees about conditions such as ASC can be beneficial. It can help affected but undiagnosed individuals gain insight into their mental state, potentially enabling them to receive the necessary assistance. Additionally, fostering a climate of understanding and empathy within the workplace can increase employees' comfort in disclosing their neurodiverse status [S9].

\subsection{Tools and Assistive Technologies}

\subsubsection{Leverage assistive technologies for employment training and task completion}
To support employment training and task completion, research has suggested the use of assistive technologies for people with ASC. Study S13 suggested designing a comprehensive virtual collaboration platform to integrate various access technologies seamlessly. This platform may include robust features such as enhanced live captioning, advanced screen readers, text read-aloud functions, and speech dictation tools, all natively built into the system to reduce the need for switching between different tools. By integrating these features into video conferencing tools, such as Zoom, Microsoft Teams, and Google Meet, the platform could provide multimodal entry and playback of text chat, which would help users manage the temporal demands of understanding content and formulating responses during meetings. Additionally, text-based collaboration tools like Slack could benefit from asynchronous multimodal interactions, enabling users to input and receive information through speech-to-text and text-to-speech functionalities. This integration would not only facilitate more accessible and efficient communication for autistic employees but also promote the normalisation and routine use of these technologies across organisations, thereby reducing the need for individuals to disclose their disabilities to request accommodations [S13].

To address the challenges autistic individuals face in interpreting complex social-emotional cues, a specialised application can be developed to translate these cues into more understandable formats [S14]. Leveraging machine learning, the application could analyse verbal and nonverbal signals during virtual communication and provide simplified summaries or visualisations. For instance, it could use bubble visualisations to represent facial expressions, making it easier for users to grasp emotional states. This tool could also offer post-conversation analysis, allowing users to reflect on social interactions and improve their understanding over time. Additionally, wearable assistive technologies, like those integrated with Google Glass, could be adapted for real-time emotion recognition, providing instant feedback during face-to-face interactions. The application could flag incongruent cues, such as mismatched body language and tone of voice, to alert users to potential misunderstandings, thus enhancing their social navigation skills. Importantly, this technology would serve as an augmentative aid, empowering autistic individuals to make informed decisions without prescribing specific actions, thereby fostering greater confidence and autonomy in social situations [S14].

Advanced applications or technology can be incorporated to design job interview training programs to further cater to the needs of autistic people or employees [S11]. This technology could integrate a comprehensive training suite that includes video modelling, virtual reality (VR) practice sessions, and real-time feedback mechanisms. The platform, accessible via both web and mobile interfaces, would offer personalised training modules tailored to different job roles and interview scenarios. Using a combination of AI-driven assessments and input from human resources experts, the application could provide detailed feedback on interview skills. Additionally, incorporating tools to measure and enhance social responsiveness would help users improve their communication and social interaction abilities. Such a platform would not only prepare autistic individuals for job interviews but also offer ongoing support to enhance workplace integration and success [S11].

Specialised software applications can be developed that emphasise clear visual design and user-friendly interfaces. This application would utilise well-designed visual aids with distinct colours and layouts to convey information effectively. Educational materials within the app would feature consistent visual design principles, ensuring clarity across written instructions, code editors, and the programming language interface. The application could include customizable visual themes to accommodate individual preferences and sensory sensitivities. Additionally, it would differentiate clearly between software tools and tutorial content, minimising confusion and enhancing learning and productivity for autistic users. Such a technology would support better engagement and comprehension, facilitating a more inclusive and supportive environment for autistic individuals in educational and professional settings [S19].

AI-powered conversational assistants can be integrated into communication and meeting platforms [S14]. This AI assistant would facilitate smoother interactions by transforming complex verbal, nonverbal, and textual cues into more comprehensible formats, reducing the cognitive load on autistic individuals. Key features could include real-time agenda updates and notifications to keep track of the meeting flow, explicit visual and auditory cues indicating the current speaker and managing turn-taking, and tools to handle interruptions and monopolisation. Additionally, the AI could provide live captioning and post-meeting summaries with appended shared content for enhanced comprehension and review. By offering these supports, the AI would help autistic employees focus on the substance of discussions and foster better relationships within the workplace [S14].

For employment interviews and training of autistic individuals or employees, a virtual reality-based interview training platform can be developed [S11]. These platforms can provide dynamic, individually tailored training for job interviews. These systems can integrate real-time stress detection, eye-gaze monitoring, and emotion recognition to adapt interview scenarios according to the participant's responses and stress levels. The platform could enhance interviewing skills by offering a flexible conversation mechanism that dynamically adjusts based on the participant's real-time feedback. A Dashboard feature for job coaches and employers can be incorporated to visualise collected data, offering insights into the participant’s stress responses, eye-gaze patterns, and emotional states during interviews. This data can inform and refine interview protocols to better accommodate autistic candidates, promoting more inclusive hiring practices. Additionally, the platform can be designed to allow participants to practice various interview scenarios, providing feedback on eye contact, speech pace, and anxiety management. The inclusion of features like a nonverbal expression tool (Whiteboard) and the ability to rephrase questions ensures that autistic individuals can better navigate and prepare for real-world interview situations. In summary, these user-centric interview training platforms can provide valuable insights to employers to foster an understanding and supportive interview environment [S11].

\subsubsection{Use CMC technologies to help increase attention and flexibility}

To better cater to the needs of autistic people or employees, Computer-mediated communication (CMC) such as video conferencing applications could incorporate several key improvements [S2, S13, S14, S18]. Firstly, these applications should offer customizable settings that allow users to filter or limit distractions, such as background noise or visual clutter. For instance, features like background blurring, already present in some platforms like Skype, should be enhanced and more widely adopted [S14]. Additionally, options to minimise the visual stimuli from other participants, such as reducing screen brightness or hiding non-speaking participants, could help maintain focus. Secondly, providing lower-bandwidth communication channels could be beneficial by reducing the pressure to maintain eye contact, which can be a significant source of stress. These channels might include text-based chat or voice-only options that allow for more comfortable and effective communication. These improvements would collectively help autistic employees manage their socio-emotional cognitive resources more efficiently and engage in more authentic and less stressful interactions [S14].

For online teaching, study S2 has recommended that educators create plain Zoom backgrounds (e.g., a light blue background with a simple logo) to reduce distractions within classrooms and for instructor identification purposes. Backgrounds can cause autistic individuals to be distracted, and in a hybrid camp, there could be many videos being displayed, causing a multitude of distractions. Educators can also create simple backgrounds and share them with hybrid students so that everyone has the same background when sharing their videos. Backgrounds could also be provided to the students to maintain consistency with the educators and reduce unnecessary distractions with their backgrounds [S2].

To improve the video conferencing application for autistic people or employees, several features could be added. First, implementing an option to keep the user's video on while asking other participants to turn theirs off can help maintain focus and reduce stress related to others' appearances or clothing. Additionally, providing the ability for users to control the video backgrounds and audio settings of other participants from their end would be highly beneficial. This means that regardless of what the background looks like on the other participants' side, users should have the capability to modify it according to their preferences, thus minimising potential distractions. Incorporating these features would create a more accommodating and less stressful virtual environment for autistic individuals [S13]. 

\begin{figure} 
\centering
\includegraphics[trim=14 125 8 10,clip, width = \textwidth]{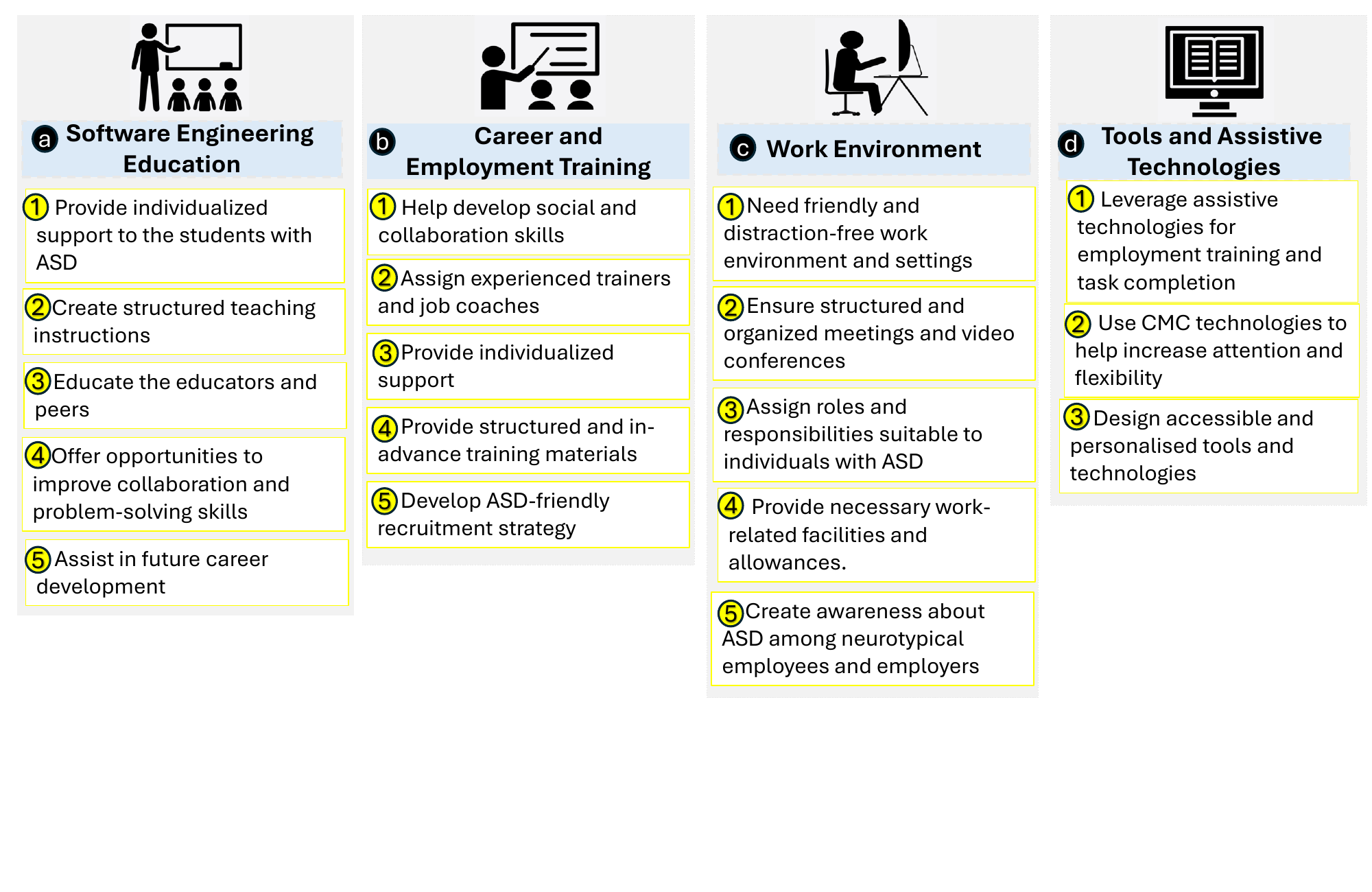}
 \caption{Summary of the findings of this study}
 \label{Findings of this study}
 \end{figure}

\subsubsection{Design accessible and personalised tools and technologies}

Designing accessible programming tools for software development and testing (e.g., API testing tools)  to cater to the needs of autistic software developers can significantly reduce the barriers they face [S26]. These tools must be thoughtfully created with accessibility as a primary focus from the initial design stages, rather than retrofitting them later. By prioritising clear, straightforward interfaces, customizable settings, and sensory-friendly designs, these tools can better accommodate the diverse needs of autistic individuals [S26].

For designing accessible programming tools, it is crucial to limit the number of programming blocks and choices available to beginners, similar to the approach used by Pocket Code and ScratchJr [S6]. This can help prevent autistic students from feeling overwhelmed and aid in maintaining their focus. Tools like Code.org, which offer project-specific constraints and templates, can also be beneficial by presenting only relevant options and media, thereby reducing cognitive load. Additionally, providing intuitive touch-based interfaces, as seen in mobile-based tools like ScratchJr and Pocket Code, can be more user-friendly for autistic students compared to traditional input devices like a mouse and keyboard. This is particularly important as these students might face difficulties with fine motor skills required for using more complex input devices. Hence, ensuring compatibility with touch gestures can enhance the overall usability of the programming tool. Conversely, tools that rely on less intuitive input methods, such as Kodu with its requirement for Xbox controllers, mice, or keyboards, might present additional challenges for these students [S6, S17, S19].

To design simplified programming activities, it is essential to simplify and streamline the coding process [S3, S5]. This can be achieved by reducing the length of programming statements, minimising the number of punctuation marks, and avoiding the use of complex identifiers that mix letters, digits, and underscores [S3]. Each block of code should be concise, ideally containing no more than five lines, and should be independently executable or verifiable to facilitate easier debugging. These simplified activities help to prevent overstimulation and reduce cognitive load, making the learning process more manageable and less stressful for autistic students. Additionally, such streamlined code is more compatible with screen readers, further enhancing accessibility and ensuring that all students can effectively engage with the programming content [S3].

Again, the programming tools need to be compatible with mobile devices, such as smartphones and tablets, to facilitate easier access and interaction, especially for those with motor difficulties [S7]. The content should be engaging and relevant to the interests of autistic students, incorporating diverse media and templates on topics like transportation and space exploration. Providing content suggestions tailored to the user’s profile can help maintain interest, particularly for those who are less motivated. Information should be presented visually, utilising icons and symbols to label objects and support text labels throughout the interface. Sounds should be optional, off by default, and with adjustable volume to accommodate sensory sensitivities. Limiting the choices of media elements and programming blocks to a manageable and relevant subset prevents overwhelming users and aids focus on the project’s goals. For example, in a “space racing game” project, limit background images to space-related scenes. Providing structured scaffolding through templates guides users through project creation steps, offering a visual framework that simplifies complex tasks. Programming elements should be available at different levels of abstraction to cater to varying abilities, such as offering high-level commands like “jump forward” for users who find low-level commands too complex [S7]. Personalisation is crucial; user profiles should store relevant personal information, and the tool should record interaction histories to adapt and evolve with the user’s needs. The interface should adjust based on these profiles, offering the appropriate level of abstraction and relevant features, while also allowing manual customisation of preferences like font size and colour. All changes should be communicated to users beforehand and implemented in small, manageable steps to avoid overwhelming them. By focusing on these design principles, programming tools can become more accessible, engaging, and supportive for autistic students, fostering an environment conducive to learning and creativity [S7].

Serious games (SG) have demonstrated significant potential in enhancing learning and motivation for typically developing children, and this potential can be harnessed to cater to the unique needs of autistic students [S16, S27]. For autistic learners, well-designed SG can provide an engaging and structured learning environment that accommodates their specific learning styles. These games can be tailored to include visual supports, clear instructions, and repetitive practice, which are beneficial for students with autism. Furthermore, SG can offer a safe space for students to practice social interactions and communication skills, which are often challenging for them. By incorporating elements that address sensory sensitivities and reduce anxiety, SG can create a more inclusive and effective educational experience for autistic students, helping them to grasp complex concepts and skills in a more enjoyable and supportive manner [S16, S27].

Additionally, it is crucial to integrate content that aligns with their interests and use engaging media on splash screens and lock screens to capture their attention [S17]. Suggesting popular content to unmotivated users can help increase engagement, provided they are open to new information. Information should be presented visually, using icons and symbols, focusing on features rather than media content or programming elements, even if the user shows fixation on them. Goal-oriented templates should scaffold towards projects that teach relevant skills such as communication and collaboration, ensuring that these projects are appropriate for children with ASC. Personalisation and customisation are key; therefore, user profiles should store personal information and preferences, which can be collected through automated tests or manually by caretakers. Recording interaction histories will allow for personalised experiences, with the tool automatically updating user records. Manual preference settings should be supported, enabling users to override automatic personalisation and choose options such as font size and colour, ensuring the tool adapts to individual needs and preferences [S17].

Another effective strategy is to constrain the number of available programming blocks to a manageable subset [S6]. This approach, exemplified by Pocket Code's use of 'beginner blocks' and Code.org's project-specific constraints, helps prevent users from feeling overwhelmed by too many choices. Similarly, ScratchJr's limitation of blocks and media within sample projects ensures that students can focus on a simplified, structured set of options. Tools should also leverage intuitive, touch-based interactions, which are more accessible than traditional input devices like a mouse or keyboard. Mobile-based platforms such as ScratchJr and Pocket Code, which use touch gestures, can be particularly beneficial. While Code.org also supports touch gestures through mobile browsers, it is optimised for laptops and desktops, which might pose challenges. In contrast, platforms like Kodu, which rely on more complex input devices such as Xbox controllers,a  mouse, or a keyboard, can overwhelm users with too many choices and interaction difficulties. Therefore, focusing on intuitive touch interactions and limiting choices can significantly enhance the accessibility of programming tools for autistic students [S6].

Programming activities should be simplified by reducing the length of statements and minimising the use of punctuation marks and mixed identifiers [S3]. This simplification helps prevent cognitive overload and makes the code easier to understand. Additionally, each block of code should be limited to no more than five lines and designed to be independently executable or verifiable. This approach simplifies debugging and makes it easier for students to focus on small, manageable tasks. Such strategies not only reduce stress for autistic students but also enhance compatibility with screen readers, making the learning process more inclusive and accessible. By implementing these adjustments, programming tools can better support the diverse needs of autistic students, fostering a more effective and comfortable learning environment [S3].

\section{Threats to Validity}
In this section, we will discuss the limitations of this study and actions taken to mitigate these. During our study selection process, some relevant papers may be overlooked. We followed a systematic approach, adhering to SLR guidelines and best practices by \citet{kitchenham2007guidelines} to mitigate this risk. For instance, we utilised Scopus, IEEE, and ACM digital libraries that index major computer science and software engineering research papers. To further minimise omissions, we employed both backward and forward snowballing to identify studies that may have been missed during the automatic search. To prevent author-related selection bias, we conducted a pilot study on a sample of 10 papers to ensure consistency in the selection process. Furthermore, any uncertainty regarding the inclusion or exclusion of a study was resolved through discussions between the first two authors before making a final decision.

To reduce potential bias in the validity of conclusions due to varying interpretations of the results \citep{ampatzoglou2019identifying}, the first author identified the codes and themes through the thematic analysis. This information was subsequently shared with all co-authors. Codebooks were revised and updated in response to suggestions and feedback provided during weekly research meetings between the first and second authors.

Despite the mitigation strategies employed in this study, we acknowledge that our reported list of success factors may not be exhaustive due to inherent internal and external biases (e.g., potential omission of primary studies and the thematic analysis process). We therefore urge readers to consider these limitations when interpreting our findings. 

\section{Conclusion}
In light of the discussed needs and the challenges faced by people with ASC in both software engineering education and employment, our study aims to investigate the factors that can maximise the employment success of people with ASC in software engineering. By conducting a Systematic Review (SR) of 30 studies, we identify 18 potential success factors categorised across 4 categories: Software Engineering Education, Career and Employment Training, Work Environment, and Tools and Assistive Technologies. Our results demonstrate that several factors can potentially improve software engineering education and career training for individuals with ASC. These include providing personalised support to students, employing structured teaching methods, educating instructors and neurotypical peers about autism, creating opportunities to develop collaboration and problem-solving skills (e.g., organising coding camps), and assisting with future career development through initiatives (e.g., career seminars and online courses). In the workplace, we identify several factors that can contribute to a more positive and productive environment for employees with ASC. These factors include creating a welcoming and distraction-free workspace, ensuring structured meetings, assigning tasks suited to the strengths of individuals with ASC, providing necessary accommodations and allowances, and fostering awareness about autism among both employers and neurotypical employees. Our study also discusses improvements required in existing tools and assistive technologies. This includes designing user interfaces for programming tools that are both simple and game-based for children with ASC, incorporating new functionalities within video conferencing technologies to better support increased focus and attention. The success factors identified in this study offer actionable insights to educators and organisations. These insights highlight the necessary modifications required within both educational settings and workplaces to effectively meet the employment needs of individuals with ASC within the software industry.

 \section*{Conflict of interest}
Not applicable.

\section*{Availability of data}
All data have been made available via an online appendix.

\section*{Code availability}
Not applicable.

\bibliographystyle{spbasic}      


\begin{thebibliography}{38}
\providecommand{\natexlab}[1]{#1}
\providecommand{\url}[1]{{#1}}
\providecommand{\urlprefix}{URL }
\expandafter\ifx\csname urlstyle\endcsname\relax
  \providecommand{\doi}[1]{DOI~\discretionary{}{}{}#1}\else
  \providecommand{\doi}{DOI~\discretionary{}{}{}\begingroup \urlstyle{rm}\Url}\fi
\providecommand{\eprint}[2][]{\url{#2}}

\bibitem[{{American Psychiatric Association}(2013)}]{APA2013}
{American Psychiatric Association} (2013) Diagnostic and Statistical Manual of Mental Disorders: DSM-5. American Psychiatric Association, Washington, DC

\bibitem[{Ampatzoglou et~al.(2019)Ampatzoglou, Bibi, Avgeriou, Verbeek, and Chatzigeorgiou}]{ampatzoglou2019identifying}
Ampatzoglou A, Bibi S, Avgeriou P, Verbeek M, Chatzigeorgiou A (2019) Identifying, categorizing and mitigating threats to validity in software engineering secondary studies. Information and Software Technology 106:201--230

\bibitem[{Annabi et~al.(2017)Annabi, Sundaresan, and Zolyomi}]{annabi2017s}
Annabi H, Sundaresan K, Zolyomi A (2017) It’s not just about attention to details: Redefining the talents autistic software developers bring to software development [conference session]. In: Hawaii International Conference on System Sciences, Honolulu, Hawaii, United States. http://hdl. handle. net/10125/41827

\bibitem[{Austin and Pisano(2017)}]{austin2017neurodiversity}
Austin RD, Pisano GP (2017) Neurodiversity as a competitive advantage. Harvard Business Review 95(3):96--103

\bibitem[{Baldwin et~al.(2014)Baldwin, Costley, and Warren}]{baldwin2014employment}
Baldwin S, Costley D, Warren A (2014) Employment activities and experiences of adults with high-functioning autism and asperger’s disorder. Journal of autism and developmental disorders 44(10):2440--2449

\bibitem[{Begel et~al.(2021)Begel, Dominic, Phillis, Beeson, and Rodeghero}]{begel2021remote}
Begel A, Dominic J, Phillis C, Beeson T, Rodeghero P (2021) How a remote video game coding camp improved autistic college students' self-efficacy in communication. In: Proceedings of the 52nd ACM Technical Symposium on Computer Science Education, pp 142--148

\bibitem[{Booth(2016)}]{booth2016autism}
Booth J (2016) Autism equality in the workplace: removing barriers and challenging discrimination. Jessica Kingsley Publishers

\bibitem[{Braun and Clarke(2006)}]{braun2006using}
Braun V, Clarke V (2006) Using thematic analysis in psychology. Qualitative research in psychology 3(2):77--101

\bibitem[{Croft et~al.(2022)Croft, Xie, and Babar}]{croft2022data}
Croft R, Xie Y, Babar MA (2022) Data preparation for software vulnerability prediction: A systematic literature review. IEEE Transactions on Software Engineering 49(3):1044--1063

\bibitem[{Du et~al.(2018)Du, Wimmer, and Rada}]{du2018hour}
Du J, Wimmer H, Rada R (2018) “hour of code”: A case study. Information Systems Education Journal 16(1):51

\bibitem[{Eiselt and Carter(2018)}]{eiselt2018integrating}
Eiselt K, Carter P (2018) Integrating social skills practice with computer programming for students on the autism spectrum. In: 2018 IEEE Frontiers in Education Conference (FIE), IEEE, pp 1--5

\bibitem[{Elshahawy et~al.(2020)Elshahawy, Bakhaty, and Sharaf}]{elshahawy2020developing}
Elshahawy M, Bakhaty M, Sharaf N (2020) Developing computational thinking for children with autism using a serious game. In: 2020 24th international conference information visualisation (IV), IEEE, pp 761--766

\bibitem[{Fan et~al.(2023)Fan, Xiong, and Sun}]{fan2023deepasdpred}
Fan Y, Xiong H, Sun G (2023) Deepasdpred: a cnn-lstm-based deep learning method for autism spectrum disorders risk rna identification. BMC bioinformatics 24(1):261

\bibitem[{Fern{\'a}ndez-L{\'o}pez et~al.(2013)Fern{\'a}ndez-L{\'o}pez, Rodr{\'\i}guez-F{\'o}rtiz, Rodr{\'\i}guez-Almendros, and Mart{\'\i}nez-Segura}]{fernandez2013mobile}
Fern{\'a}ndez-L{\'o}pez {\'A}, Rodr{\'\i}guez-F{\'o}rtiz MJ, Rodr{\'\i}guez-Almendros ML, Mart{\'\i}nez-Segura MJ (2013) Mobile learning technology based on ios devices to support students with special education needs. Computers \& Education 61:77--90

\bibitem[{Gribble et~al.(2017)Gribble, Hansen, Harlow, and Franklin}]{gribble2017cracking}
Gribble J, Hansen A, Harlow D, Franklin D (2017) Cracking the code: the impact of computer coding on the interactions of a child with autism. In: Proceedings of the 2017 conference on interaction design and children, pp 445--450

\bibitem[{Haanappel and Brinkkemper(2010)}]{haanappel2010software}
Haanappel S, Brinkkemper S (2010) Software testing by people with autism. In: International Conference on Computer Safety, Reliability, and Security, Springer, pp 251--262

\bibitem[{Hayward et~al.(2019)Hayward, McVilly, and Stokes}]{hayward2019autism}
Hayward SM, McVilly KR, Stokes MA (2019) Autism and employment: What works. Research in Autism Spectrum Disorders 60:48--58, \doi{10.1016/j.rasd.2019.01.006}, \urlprefix\url{https://doi.org/10.1016/j.rasd.2019.01.006}

\bibitem[{Hedley et~al.(2018)Hedley, Cai, Uljarevic, Wilmot, Spoor, Richdale, and Dissanayake}]{hedley2018transition}
Hedley D, Cai R, Uljarevic M, Wilmot M, Spoor JR, Richdale A, Dissanayake C (2018) Transition to work: Perspectives from the autism spectrum. Autism 22(5):528--541

\bibitem[{Kitchenham and Charters(2007)}]{kitchenham2007guidelines}
Kitchenham B, Charters S (2007) Guidelines for performing systematic literature reviews in software engineering. Technical report, EBSE Technical Report EBSE-2007-01

\bibitem[{Ko and Davis(2017)}]{ko2017computing}
Ko AJ, Davis K (2017) Computing mentorship in a software boomtown: Relationships to adolescent interest and beliefs. In: Proceedings of the 2017 ACM conference on international computing education research, pp 236--244

\bibitem[{Lin et~al.(2020)Lin, Wen, Han, Zhang, and Xiang}]{lin2020software}
Lin G, Wen S, Han QL, Zhang J, Xiang Y (2020) Software vulnerability detection using deep neural networks: a survey. Proceedings of the IEEE 108(10):1825--1848

\bibitem[{M{\'a}rquez et~al.(2024)M{\'a}rquez, Pacheco, Astudillo, Taramasco, and Calvo}]{marquez2024inclusion}
M{\'a}rquez G, Pacheco M, Astudillo H, Taramasco C, Calvo E (2024) Inclusion of individuals with autism spectrum disorder in software engineering. Information and Software Technology p 107434

\bibitem[{Morris et~al.(2015)Morris, Begel, and Wiedermann}]{morris2015understanding}
Morris MR, Begel A, Wiedermann B (2015) Understanding the challenges faced by neurodiverse software engineering employees: Towards a more inclusive and productive technical workforce. In: Proceedings of the 17th International ACM SIGACCESS Conference on computers \& accessibility, pp 173--184

\bibitem[{Moster et~al.(2022)Moster, Kokinda, Re, Dominic, Lehmann, Begel, and Rodeghero}]{moster2022can}
Moster M, Kokinda E, Re M, Dominic J, Lehmann J, Begel A, Rodeghero P (2022) " can you help me?" an experience report of teamwork in a game coding camp for autistic high school students. In: Proceedings of the ACM/IEEE 44th International Conference on Software Engineering: Software Engineering Education and Training, pp 50--61

\bibitem[{Nicholas et~al.(2017)Nicholas, Hodgetts, Zwaigenbaum, Smith, Shattuck, Parr, Conlon, Germani, Mitchell, Sacrey et~al.}]{nicholas2017research}
Nicholas DB, Hodgetts S, Zwaigenbaum L, Smith LE, Shattuck P, Parr JR, Conlon O, Germani T, Mitchell W, Sacrey L, et~al. (2017) Research needs and priorities for transition and employment in autism: Considerations reflected in a “special interest group” at the international meeting for autism research. Autism Research 10(1):15--24

\bibitem[{Ribu(2010)}]{ribu2010teaching}
Ribu K (2010) Teaching computer science to students with asperger’s syndrome. Tapir Akademisk Forlag

\bibitem[{Richards(2012)}]{richards2012examining}
Richards J (2012) Examining the exclusion of employees with asperger syndrome from the workplace. Personnel Review 41(5):630--646

\bibitem[{Roux(2015)}]{roux2015national}
Roux AM (2015) National autism indicators report: Transition into young adulthood. AJ Drexel Autism Institute

\bibitem[{Sarker et~al.(2024)Sarker, Jayatilaka, Haggag, Liu, and Babar}]{sarker2024multi}
Sarker O, Jayatilaka A, Haggag S, Liu C, Babar MA (2024) A multi-vocal literature review on challenges and critical success factors of phishing education, training and awareness. Journal of Systems and Software 208:111899

\bibitem[{Sbaraini et~al.(2011)Sbaraini, Carter, Evans, and Blinkhorn}]{sbaraini2011grounded}
Sbaraini A, Carter SM, Evans RW, Blinkhorn A (2011) How to do a grounded theory study: a worked example of a study of dental practices. BMC medical research methodology 11(1):1--10

\bibitem[{Shattuck et~al.(2012)Shattuck, Narendorf, Cooper, Sterzing, Wagner, and Taylor}]{shattuck2012postsecondary}
Shattuck PT, Narendorf SC, Cooper B, Sterzing PR, Wagner M, Taylor JL (2012) Postsecondary education and employment among youth with an autism spectrum disorder. Pediatrics 129(6):1042--1049

\bibitem[{Stuurman et~al.(2019)Stuurman, Passier, Geven, and Barendsen}]{stuurman2019autism}
Stuurman S, Passier HJ, Geven F, Barendsen E (2019) Autism: Implications for inclusive education with respect to software engineering. In: Proceedings of the 8th Computer Science Education Research Conference, pp 15--25

\bibitem[{Tang(2021)}]{tang2021understanding}
Tang J (2021) Understanding the telework experience of people with disabilities. Proceedings of the ACM on Human-Computer Interaction 5(CSCW1):1--27

\bibitem[{Taylor et~al.(2019)Taylor, Smith~DaWalt, Marvin, Law, and Lipkin}]{taylor2019sex}
Taylor JL, Smith~DaWalt L, Marvin AR, Law JK, Lipkin P (2019) Sex differences in employment and supports for adults with autism spectrum disorder. Autism 23(7):1711--1719

\bibitem[{Wehman et~al.(2017)Wehman, Schall, McDonough, Graham, Brooke, Riehle, Brooke, Ham, Lau, Allen et~al.}]{wehman2017effects}
Wehman P, Schall CM, McDonough J, Graham C, Brooke V, Riehle JE, Brooke A, Ham W, Lau S, Allen J, et~al. (2017) Effects of an employer-based intervention on employment outcomes for youth with significant support needs due to autism. Autism 21(3):276--290

\bibitem[{Wei et~al.(2014)Wei, Christiano, Yu, Blackorby, Shattuck, and Newman}]{Wei2014}
Wei X, Christiano ERA, Yu JW, Blackorby J, Shattuck P, Newman LA (2014) Postsecondary pathways and persistence for stem versus non-stem majors: Among college students with an autism spectrum disorder. Journal of Autism and Developmental Disorders 44(5):1159--1167, \doi{10.1007/s10803-013-1978-5}, \urlprefix\url{https://doi.org/10.1007/s10803-013-1978-5}

\bibitem[{Wohlin(2014)}]{wohlin2014guidelines}
Wohlin C (2014) Guidelines for snowballing in systematic literature studies and a replication in software engineering. In: Proceedings of the 18th international conference on evaluation and assessment in software engineering, pp 1--10

\bibitem[{Zubair et~al.(2023)Zubair, Brown, Hughes-Roberts, and Bates}]{zubair2023designing}
Zubair MS, Brown DJ, Hughes-Roberts T, Bates M (2023) Designing accessible visual programming tools for children with autism spectrum condition. Universal Access in the Information Society 22(2):277--296

\end{thebibliography}

%
%

\end{document}